\documentclass[preprint,12pt, a4paper]{elsarticle}

\usepackage{amsthm,amsmath,amssymb,amsfonts}%
\usepackage{tabularx}
\usepackage{caption}
\usepackage{subcaption}
\usepackage{float}
\usepackage{multicol}
\usepackage{multirow}
\usepackage{booktabs}
\usepackage{caption}
\usepackage[utf8]{inputenc}
\usepackage{soul}
\usepackage{listings}%
\usepackage{threeparttable}
\usepackage{textcomp}%
\usepackage{xcolor}%
\usepackage{mathrsfs}%
\usepackage[polish,english]{babel} 
\usepackage{setspace}
\usepackage{siunitx}

\journal{Acta Materialia}

\begin{document}
	\emergencystretch 3em
	
	\begin{frontmatter}
		
		\title{Dominant-pair free energies predict phase selection in high-entropy alloys}
		
		\author[a]{Dennis Boakye}
		\author[a]{Chuang Deng \corref{b}}
		\cortext[b]{Corresponding author}
		\ead{chuang.deng@umanitoba.ca}
		
		\affiliation[a]{organization={Mechanical Engineering, University of Manitoba},
			addressline={66 Chancellors Cir}, 
			city={Winnipeg},
			postcode={R3T 2N2}, 
			state={Manitoba},
			country={Canada}}
		
		\begin{abstract}
		Phase selection in multicomponent alloys is governed by the competition between entropic stabilization of disordered solutions and enthalpic driving forces for chemical ordering. However, widely used parametric criteria reduce it to a single scalar, carrying no explicit free energy for any competing ordered phase. Herein, we develop a thermodynamic framework based on the semi-empirical macroscopic atom model and the Dinsdale lattice stability database to fill this gap. We show that a dominant-pair mechanism, in which the Al-transition-metal interaction family dominates the ordering enthalpy, enables the complex multicomponent B2-ordering problem to be reduced to an effective pseudo-binary system with an analytically evaluated Bragg-Williams free energy. Combined with a minimum-free-energy classifier, the framework predicts the lowest-energy phase as a function of composition and temperature. This provides continuous phase stability maps rather than the single-value predictions of conventional descriptors. Demonstrated on high-entropy alloys using a dataset of 269 experimentally characterized samples, the model outperforms widely used phase-selection criteria in the class-balanced macro-F1 metric and achieves 77.9\% on the well-posed three-class task, outperforming the valence electron concentration criterion. The model is general by construction and computationally efficient for predicting phase stability in multicomponent alloys over a broad range of compositions and temperatures.
		\end{abstract}
		
		
	\end{frontmatter}
	
	\section{Introduction}
	
	The central proposition of high-entropy alloys (HEAs), pioneered independently by Yeh et al.~\cite{yeh2004nanostructured} and Cantor et al.~\cite{cantor2004microstructural}, is that near-equimolar mixtures of five or more principal elements can crystallize as single-phase solid solutions stabilized by their large configurational entropy. The classic CoCrFeMnNi alloy forms a single-phase FCC solid solution with outstanding cryogenic toughness~\cite{gludovatz2014fracture}, while refractory HEAs such as BCC NbMoTaW retain solid-solution stability and strength above \SI{1600}{\celsius}~\cite{senkov2010refractory,senkov2011mechanical}. Yet it emerged early that configurational entropy is neither sufficient nor always necessary for single-phase formation~\cite{otto2013relative,sheng2011phase}. Phase selection reflects a competition between entropic stabilization of the disordered solution and enthalpic driving forces toward ordered phases, and this balance shifts sharply with composition. Adding Al to the Cantor base, for instance, drives a progression from FCC at low Al, through a duplex FCC$+$BCC/B2 regime, to a BCC/B2-dominated microstructure at high Al~\cite{he2016precipitation,wang2012effects,kumar2018effect,hsu2019effects}, because Al acts at once as a BCC stabilizer and a B2-ordering promoter.
	
	Faced with a composition space that is too vast to be exhaustively explored, parametric criteria have been developed that triage candidates from tabulated elemental data. Zhang et al.~\cite{zhang2008solid} identified the atomic-size mismatch $\delta$ and mixing enthalpy $\Delta H$ as controlling variables. Later, Yang and Zhang~\cite{yang2012prediction} combined them into $\Omega = T_m \Delta S / |\Delta H|$, requiring $\Omega \geq 1.1$ and $\delta \leq 6.6\%$ for a solid solution. Guo et al.~\cite{guo2011effect} also introduced the valence-electron-concentration (VEC) criterion, separating FCC ($\mathrm{VEC} \geq 8$) from BCC ($\mathrm{VEC} \leq 6.87$) stability, while Sheng and Liu~\cite{sheng2011phase} delimited a solid-solution region in $\Delta H$--$\delta$ space. In the early growth years of HEA studies, Senkov and Miracle~\cite{senkov2016new} proposed a temperature-dependent parameter $k_1^{\mathrm{cr}}(T)$ accounting for competing intermetallic formation. With configurational entropy at the heart of HEAs, Ye et al.~\cite{ye2015design} condensed the competition into $\phi = (S_C - S_H)/|S_E|$, with $S_E$ an excess packing entropy from Mansoori hard-sphere theory~\cite{mansoori1971equilibrium} and a critical value $\phi_c \approx 20$. Moreover, electronic and density-functional screening descriptors have extended the toolkit further~\cite{troparevsky2015criteria}.
	
	These criteria are useful but limited in two structural ways that motivate this work. First, each compresses a multidimensional free-energy landscape onto a single scalar with a fitted threshold, and such a reduction cannot perform the two distinct discrimination phase selection demands. Distinguishing FCC from BCC is a question of lattice and band filling, whereas deciding whether an ordered intermetallic competes is a free-energy comparison against a specific phase. A ratio such as $\Omega$ carries no lattice information, and a descriptor such as VEC carries none about intermetallic competition making each blind to one axis. Second, the criteria have not, to our knowledge, been benchmarked against one another and against an explicit free-energy calculation on a single large dataset under a held-out protocol, so their relative accuracy and the headroom for a more complete model remain unquantified.
	
	The semi-empirical macroscopic atom model (MAM) \cite{miedema1980cohesion,boer1988cohesion} offers a route to the underlying energetics without fitted binary parameters. Using three tabulated elemental quantities, the model predicts the dilute-limit interaction enthalpies of any binary pair, and hence the mixing enthalpy of a multicomponent solution, with the Takeuchi--Inoue tabulation~\cite{takeuchi2005classification} providing a validation benchmark. What the MAM lacks is precisely the two ingredients missing from the scalar criteria. First, the model carries no crystal-structure information because the chemical enthalpy is structure-independent. Second, it constructs no competing ordered-phase free energy. Supplying a structural energy that resolves FCC from BCC and an explicit ordered-phase free energy converts the MAM energetics into an actual phase prediction, and doing so from tabulated data alone avoids the fitted databases required by CALPHAD.
	
	In this work, we construct such a model and benchmark it fairly, providing a comparison currently lacking in the literature. The classifier assigns an explicit Gibbs free energy to every candidate phase, drawing on a complete Miedema interaction database, tabulated unary lattice stabilities, and a dominant-pair treatment of B2 ordering. The model then selects the most stable phase by direct comparison. Its distinctive ingredient is the explicit ordered-phase free energy, which the scalar criteria lack, made tractable by a dominant-pair mechanism in which a single strong interaction family supplies most of the mixing enthalpy and collapses the multicomponent ordering problem onto an effective pseudo-binary. We use this construction to deliver the first held-out comparison against the five established criteria and to expose a resolution ceiling intrinsic to how the experimental literature labels multiphase alloys, which bounds every method. The framework is general by construction and ranks the free energy of every candidate phase at any composition and temperature. While we demonstrate and validate it here on HEAs, the most stringent test available, it applies without modification to any multicomponent alloy, extending its reach well beyond the high-entropy regime that motivates it.
	
	\section{Theoretical framework}
	
The model rests on the principle that the phase an alloy adopts at temperature $T$ is the competitor with the lowest Gibbs free energy. This free energy is assembled from a chemical interaction enthalpy, a structural energy distinguishing the lattices, an entropy beyond the ideal configurational term, and, for ordered phases, an ordering free energy. We develop these contributions in the order in which they enter the classifier, beginning with the chemical enthalpy common to every state.
	
	For atomic fractions $\mathbf{c} = (c_1, \dots, c_n)$, the MAM supplies the dilute-limit enthalpies $\Delta H^{\circ}_{i\,\text{in}\,j}$ (a trace of $i$ in a matrix of $j$) and its asymmetric version $\Delta H^{\circ}_{j\,\text{in}\,i}$. The two differ because embedding an $i$ atom in $j$ costs a different energy than the reverse though some have equal energy. Instead of the symmetric average, which discards the volume dependence of this asymmetry, we use the MAM's proper weighting by surface concentrations, which account for the difference in volume of the atoms. Given a pair $\{i,j\}$ the surface fraction of $j$ is
	\begin{equation}
		x_j^{s} = \frac{c_j\, V_j^{2/3}}{c_i\, V_i^{2/3} + c_j\, V_j^{2/3}}, \qquad x_i^{s} = 1 - x_j^{s},
		\label{eq:surface}
	\end{equation}
	with $V_i$ the molar volume of element $i$, so that a larger atom presents a correspondingly larger share of the contact surface than its mole fraction alone would imply. The composition-dependent effective interaction parameter for the pair is then the surface-weighted combination of the two dilute limits,
	\begin{equation}
		\Omega_{ij}^{\mathrm{eff}}(\mathbf{c}) = x_j^{s}\,\Delta H^{\circ}_{i\,\text{in}\,j} + x_i^{s}\,\Delta H^{\circ}_{j\,\text{in}\,i},
		\label{eq:omegaeff}
	\end{equation}
	and the total chemical mixing enthalpy of the disordered solution follows from summing the pairwise contributions weighted by the product of mole fractions,
	\begin{equation}
		\Delta H_{\mathrm{chem}}(\mathbf{c}) = \sum_{i<j} \Omega_{ij}^{\mathrm{eff}}(\mathbf{c})\, c_i\, c_j.
		\label{eq:dHchem}
	\end{equation}
	Equation \eqref{eq:dHchem} generalizes the symmetric regular-solution form rather than departing from it. For equal molar volumes, Equation~\eqref{eq:surface} gives $x_i^{s}=x_j^{s}=\tfrac{1}{2}$ and Equation~\eqref{eq:omegaeff} reduces to the symmetric average, so the volume-matched transition metal (TM) pairs are treated conventionally and the correction acts only on the volume-mismatched Al--TM pairs, where the enthalpy is dominated. This chemical enthalpy is structure-independent and reflects only charge transfer and electron-density mismatch, not the lattice itself, and thus forms a baseline shared by every candidate phase.
	
To convert the chemical enthalpy into structure-resolved energies, the enthalpy of the solid solution on a given crystal structure $\varphi$ is obtained by adding to the chemical term the composition-weighted unary lattice stabilities, standard quantities that express the energy of each pure element in a structure other than its ground state,
	\begin{equation}
		\Delta H_{\mathrm{SS}}^{\varphi}(\mathbf{c}) = \Delta H_{\mathrm{chem}}(\mathbf{c}) + \sum_i c_i\, {}^{\circ}G_i^{\,\varphi-\mathrm{FCC}},
		\label{eq:dHSS}
	\end{equation}
	where ${}^{\circ}G_i^{\,\varphi-\mathrm{FCC}}$ is the lattice stability of element $i$ in structure $\varphi$ relative to the FCC reference, taken from standard thermochemical compilations and evaluated for the elevated-temperature regime in which the alloys equilibrate  \cite{dinsdale1991sgte}. The entire difference between the FCC and BCC solid-solution enthalpies is then carried by the composition-weighted difference of lattice stabilities, $\sum_i c_i\,{}^{\circ}G_i^{\,\mathrm{BCC-FCC}}$, a quantity that is positive for the FCC-forming late TMs and Al and negative for the BCC-forming early and refractory TMs. This structural term in principle provides a first-principles-informed discriminator between FCC and BCC stability that would supersede the empirical VEC threshold. In practice, the linear rule-of-mixtures over pure-element lattice stabilities does not reproduce the experimental FCC/BCC/duplex ordering as cleanly as the electronic VEC coordinate does, because the collective $d$-band-filling physics that governs structural preference in concentrated TM solutions is not captured by a linear sum of pure-element penalties. We therefore make the deliberate, evidence-based decision to retain VEC as the structural discriminator while reserving the free-energy machinery for the solid-solution-versus-intermetallic decision, where it provides genuine and quantifiable improvement. The VEC itself is the composition-weighted mean of the elemental valence-electron counts, $\mathrm{VEC}(\mathbf{c}) = \sum_i c_i\,(\mathrm{VEC})_i$, and the established phase boundaries place the alloy in the FCC field for $\mathrm{VEC} \geq 8$, in the BCC field for $\mathrm{VEC} \leq 6.87$, and in the intervening duplex region otherwise.
	
	The free energy of each solid-solution branch requires an entropy, and here we extend the treatment beyond the ideal configurational term that the parametric criteria employ. The configurational contribution is the usual ideal-mixing expression $S_{\mathrm{conf}}(\mathbf{c}) = -R \sum_i c_i \ln c_i$, which attains its maximum at the equimolar composition and supplies the entropic stabilization that underlies the entire high-entropy concept. To this we add a magnetic contribution, because several of the $3d$ constituents such as Fe, Co, Ni, Mn, and Cr carry atomic magnetic moments whose disordering at the elevated temperatures of interest contributes a real and frequently neglected entropy \cite{schneeweiss2017magnetic}. In the paramagnetic, fully disordered limit appropriate to processing and annealing temperatures above the magnetic ordering temperatures, the magnetic entropy takes the Hillert--Jarl form
	\begin{equation}
		S_{\mathrm{mag}}(\mathbf{c}) = R \sum_i c_i \ln\!\big(\beta_i + 1\big),
		\label{eq:Smag}
	\end{equation}
	in which $\beta_i$ is the atomic magnetic moment of element $i$ in Bohr magnetons and the term vanishes for the non-magnetic elements, for which $\beta_i = 0$. The approximation inherent in Equation~\eqref{eq:Smag} is that the moments are taken at their elemental values and their modification on alloying is neglected, a simplification that is defensible at high temperature and that is in any case absent altogether from the competing criteria. The Gibbs free energy of the solid solution on structure $\varphi$ is then written as
	\begin{equation}
		G_{\mathrm{SS}}^{\varphi}(\mathbf{c},T) = \Delta H_{\mathrm{SS}}^{\varphi}(\mathbf{c}) - T\big[\,S_{\mathrm{conf}}(\mathbf{c}) + S_{\mathrm{mag}}(\mathbf{c})\,\big],
		\label{eq:GSS}
	\end{equation}
	and the same magnetic and configurational entropies are carried, with the appropriate modification described below, into the ordered competitors, so that the comparison between disordered and ordered states is made on a consistent entropic footing.
	
	The ingredient missing from scalar criteria is an explicit free energy for the competing ordered phase. In five- or six-component alloys, constructing such a free energy would be intractable without a simplification inherent in the energetics. In Al-bearing alloys, the chemical enthalpy of Equation~\eqref{eq:dHchem} is overwhelmingly dominated by the Al--TM interactions. Because Al has both a low adjusted electronegativity and a low electron density, these interactions are an order of magnitude stronger than those among the TMs, so a single pair family accounts for most of the enthalpy above a few atomic percent Al. This lets the multicomponent B2-ordering problem reduce to an effective pseudo-binary between Al, $x \equiv c_{\mathrm{Al}}$, and a composite TM species $M$ at fraction $x_B = 1 - x$, with effective interaction $\Omega_{\mathrm{RS}}$ equal to the composition-weighted Al--TM parameter from Equation~\eqref{eq:omegaeff}. The B2 (CsCl-type) structure splits the BCC lattice into two interpenetrating sublattices $\alpha$ and $\beta$, between which the species redistribute on ordering. With a long-range-order parameter $\eta$, the Al occupancies are
	\begin{equation}
		y_{\mathrm{Al}}^{\alpha} = x + \eta\, x_B, \qquad y_{\mathrm{Al}}^{\beta} = x - \eta\, x_B,
		\label{eq:occ}
	\end{equation}
	so that $\eta = 0$ recovers the disordered solution with equal occupancy of both sublattices and $\eta = x/x_B$ corresponds to complete order with all Al confined to the $\alpha$ sublattice, the TM occupancies following as $y_{M}^{p} = 1 - y_{\mathrm{Al}}^{p}$ on each sublattice $p$. Counting the Al--TM nearest-neighbor bonds across the two sublattices in the Bragg--Williams mean-field approximation yields a mixing enthalpy that is lowered quadratically by ordering,
	\begin{equation}
		\Delta H^{\mathrm{B2}}(x,\eta) = \Omega_{\mathrm{RS}}\,\big(x\, x_B + \eta^2 x_B^2\big), \qquad \Omega_{\mathrm{RS}} < 0,
		\label{eq:HB2}
	\end{equation}
	in which the first term reproduces the disordered enthalpy at $\eta = 0$ and the second, with $\Omega_{\mathrm{RS}}$ negative for the attractive Al--TM interaction, expresses the additional enthalpic stabilization gained by concentrating favorable unlike-neighbor bonds on ordering. This enthalpic gain is opposed by a loss of configurational entropy, which on the two-sublattice B2 structure takes the form
	\begin{equation}
		S^{\mathrm{B2}}(x,\eta) = -\frac{R}{2}\sum_{p\in\{\alpha,\beta\}}\Big[\,y_{\mathrm{Al}}^{p}\ln y_{\mathrm{Al}}^{p} + y_{M}^{p}\ln y_{M}^{p}\,\Big],
		\label{eq:SB2}
	\end{equation}
	reducing from the ideal value at $\eta = 0$ toward zero as full order is approached. The equilibrium degree of order at temperature $T$ minimizes the ordering free energy, and the B2 free energy is accordingly written as the disordered-BCC solid-solution energy plus the ordering correction evaluated at this minimum,
	\begin{equation}
		G^{\mathrm{B2}}(x,T) = G_{\mathrm{SS}}^{\mathrm{BCC}}(\mathbf{c},T) + \min_{\eta}\Big[\,\Omega_{\mathrm{RS}}\,\eta^2 x_B^2 - T\big(S^{\mathrm{B2}}(x,\eta) - S^{\mathrm{B2}}(x,0)\big)\Big].
		\label{eq:GB2}
	\end{equation}
	A useful consequence of this construction is that it self-regulates. If the bracketed term is differentiated twice with respect to $\eta$ at $\eta = 0$, it shows that a non-trivial ordered minimum appears only below a critical temperature
	\begin{equation}
		T_c(x) = -\frac{2\,\Omega_{\mathrm{RS}}\, x\, x_B}{R},
		\label{eq:Tc}
	\end{equation}
	so that when the dominant interaction is weak or absent such as in the Al-free refractory systems, where $\Omega_{\mathrm{RS}}$ is small the critical temperature falls below the evaluation temperature, the equilibrium order parameter is zero, and the B2 free energy collapses to that of the disordered BCC solid solution. The ordered competitor is thus automatically inactive where ordering is not physically expected, and the strength of the dominant pair, through Equations~\eqref{eq:HB2} and~\eqref{eq:Tc}, sets both the depth of the ordering stabilization and the temperature below which it operates.
	
For the refractory family, the relevant ordered competitor is the size-driven Laves phase, which we represent using an approximate two-sublattice descriptor that avoids the density-functional or CALPHAD end-member energies required for a rigorous treatment. The constituents are divided by size onto the Laves sublattices where larger atoms on the $A$ sublattice (one third of sites), smaller atoms on the $B$ sublattice (two thirds), following the $AB_2$ stoichiometry and the descriptor's free energy combines an enhanced cross-sublattice bonding enthalpy, a geometric stabilization peaked at the ideal radius ratio, and the reduced entropy of the ordered compound. Resting on estimated rather than first-principles energies, it serves to flag the size-driven intermetallic tendency of the refractory alloys rather than as an exact free energy.
	
With the four candidate free energies in hand, the classifier assigns a phase by selecting the state with the lowest free energy,
	\begin{equation}
		\text{phase}(\mathbf{c},T) = \arg\min\big\{\,G_{\mathrm{SS}}^{\mathrm{FCC}},\; G_{\mathrm{SS}}^{\mathrm{BCC}},\; G^{\mathrm{B2}},\; G^{\mathrm{Laves}}\,\big\},
		\label{eq:argmin}
	\end{equation}
	augmented by two physically motivated refinements that translate the energetic comparison into the four experimental labels. The first refinement recognizes that the FCC$+$BCC duplex microstructure corresponds physically not to a single winning phase but to the near-degeneracy of the FCC and BCC-family solid solutions, and assigns the duplex label whenever the FCC free energy and the lower of the BCC and B2 free energies lie within a small tolerance $\tau_{\mathrm{duplex}}$ of one another, the alloy then being predicted to lie on or near a two-phase tie-line. The second refinement recognizes that a strongly ordered B2 phase, as opposed to a weakly ordered one that is better described as an ordered solid solution, is reported experimentally as an intermetallic constituent, and assigns the intermetallic label when the winning B2 phase has an equilibrium order parameter exceeding a threshold $\eta_{\mathrm{IM}}$, or when the Laves descriptor is clearly lowest in energy. Within the solid-solution branches, the structural assignment between FCC and BCC is made based on the electronic VEC coordinate for the reasons discussed above. The two thresholds $\tau_{\mathrm{duplex}}$ and $\eta_{\mathrm{IM}}$ are the only adjustable quantities in the entire framework, and they are fitted exclusively on training data within the cross-validation protocol described, never on the held-out alloys against which performance is reported.
	
	To benchmark this classifier, the parametric criteria are expressed in a common form, and their definitions are summarized here to ensure an unambiguous comparison. The Yang--Zhang parameter is $\Omega = T_m \Delta S / |\Delta H|$ with the mean melting temperature $T_m = \sum_i c_i T_{m,i}$, and its associated criterion classifies an alloy as a solid solution when $\Omega \geq 1.1$ and the atomic-size mismatch $\delta = 100\sqrt{\sum_i c_i (1 - r_i/\bar{r})^2}$, with $\bar{r} = \sum_i c_i r_i$ the mean atomic radius, satisfies $\delta \leq 6.6\%$; alloys failing either condition are assigned to the intermetallic class, and those passing both are resolved into FCC, BCC, or duplex by the VEC boundaries. The Guo--Liu criterion places an alloy in the solid-solution-forming region when $\delta \leq 8.5\%$ and $-22 \leq \Delta H \leq 7$~\si{kJ/mol}, again resolving the structure by VEC within that region and otherwise predicting intermetallic or amorphous formation. The Senkov--Miracle parameter $k_1^{\mathrm{cr}}(T) = T\,\Delta S/(|\Delta H|)(1-k_2) + 1$, with $k_2 = 0.6$ and the entropy and enthalpy in consistent units, is evaluated at the characterization temperature where available and otherwise at the mean melting temperature, and an alloy is classed as a solid solution when $k_1^{\mathrm{cr}}$ exceeds the value of approximately two that marks the intermetallic boundary. The Ye parameter $\phi = (S_C - S_H)/|S_E|$ combines the configurational entropy $S_C$, the enthalpy-derived term $S_H = |\Delta H|/T_m$, and the excess packing entropy $S_E$, the last computed in the Mansoori leading-order form $S_E = -\tfrac{3}{2} R\,\eta_{\mathrm{pack}}/(1-\eta_{\mathrm{pack}})\sum_i c_i(1 - r_i/\bar{r})^2$ and averaged over the FCC and BCC packing fractions $\eta_{\mathrm{pack}} = 0.7405$ and $0.6802$; an alloy is predicted to form a single-phase solid solution when $\phi \geq 20$. Each criterion thus reduces to a rule that emits one of the four phase labels, and all are evaluated on the identical dataset and compared by the same metrics.
	
	\section{Computational methodology}
	
	The elemental data underpinning the model comprise the Miedema parameters, molar volumes, melting temperatures, valence-electron counts, atomic radii, and magnetic moments for the fourteen elements spanning the $3d$-TM plus Al family (Al, Co, Cr, Fe, Mn, Ni) and the refractory family (Ti, Zr, Hf, V, Nb, Ta, Mo, W), together with the additional minority elements (Cu, Si, Pt, Y) appearing in a small number of dataset alloys. The $\Delta H^{\circ}_{i\,\text{in}\,j}$ for all required binary pairs are drawn from a complete Miedema database (Supplemental sheet Table S1), and the unary BCC--FCC lattice stabilities are taken from the SGTE unary database \cite{dinsdale1991sgte}, evaluated at \SI{1273}{K}. The chemical enthalpy, structural energies, entropies, and ordered-phase free energies are computed exactly as set out in the preceding section, and the classifier of Equation~\eqref{eq:argmin} with its two refinements is applied to each alloy at its reported characterization temperature, defaulting to a representative annealing temperature of \SI{1273}{K} where no temperature is recorded.
	
	The validation dataset was meticulously prepared from the experimental HEA literature, spanning both the $3d$+Al family (Supplemental sheet Table S2). This includes the densely sampled Al$_x$CoCrFeMnNi and Al$_x$CoCrFeNi Al-addition series that traverse all three solid-solution regimes~\cite{kumar2018effect,hsu2019effects,he2016precipitation,wang2012effects} and the refractory family, from the NbMoTaW-type alloys~\cite{senkov2010refractory,senkov2011mechanical} to lighter and Al-modified refractory systems and combinatorial libraries~\cite{lin2015effect,zhou2023ultra,senkov2015accelerated}, many of which are reported as a BCC matrix carrying a minor secondary Laves or ordered phase~\cite{whitfield2023rate,wen2021effects}. Each entry records a composition, the reported phase constitution, the processing state, the characterization temperature where reported, and the source. Compositions appearing in molar-ratio, atomic-percent, or weight-percent conventions were converted to atomic fractions. Phase labels were assigned from the authors' stated conclusions, which rest principally on X-ray diffraction and are corroborated where available by electron microscopy and energy-dispersive spectroscopy; a secondary phase present below approximately five volume percent was not permitted to alter the primary label, so that an alloy reported as a solid-solution matrix carrying a trace intermetallic retains its solid-solution label, while an alloy in which an intermetallic is a major constituent is labeled intermetallic or multiphase. Of the $275$ alloys curated, six whose primary phase is hexagonal close-packed were flagged as out of scope for a model that resolves only the cubic FCC, BCC, and B2 structures and the Laves intermetallic, leaving $269$ in-scope alloys. The resulting class distribution is well balanced for a four-class problem, comprising $93$ BCC, $64$ FCC, $60$ duplex FCC$+$BCC, and $52$ intermetallic or multiphase alloys, and every binary pair required to evaluate any alloy is present in the Miedema database.
	
	The classifier is benchmarked against the five parametric criteria defined earlier, each implemented with its published formula and threshold. The Ye $\phi$-parameter implementation was verified by reproducing previously reported values for reference alloys, returning $\phi \approx 49$ for the equimolar TiZrHfMoCr refractory alloy in agreement with the established value, which confirms that the baseline reflects the criterion as its authors defined it rather than a weakened approximation. Performance is evaluated by stratified five-fold cross-validation where the alloys are partitioned into five folds, the two classifier thresholds are fitted by exhaustive grid search on the four training folds of each split, and predictions are recorded on the held-out fold, so that no alloy contributes to both the fitting and the scoring of the thresholds that label it. The parametric criteria carry no adjustable thresholds and are evaluated directly on all alloys. Two metrics are reported: the overall accuracy, defined as the fraction of alloys correctly classified, and the macro-averaged F1 score, defined as the unweighted mean of the per-class F1 scores. The macro-F1 is emphasized because the four classes are of unequal size and, more importantly, of unequal difficulty. A method that performs well on the easy solid-solution classes while neglecting the intermetallic class can attain a deceptively high accuracy, whereas the macro-F1, by weighting all four classes equally, rewards balanced predictive power and penalizes the neglect of any single class. All computations were carried out in Python.
	
	\section{Results and discussion}
	
	\subsection{Validation of the energetic backbone}
	
\begin{figure}[!ht]
		\centering
		\includegraphics[width=0.7\linewidth]{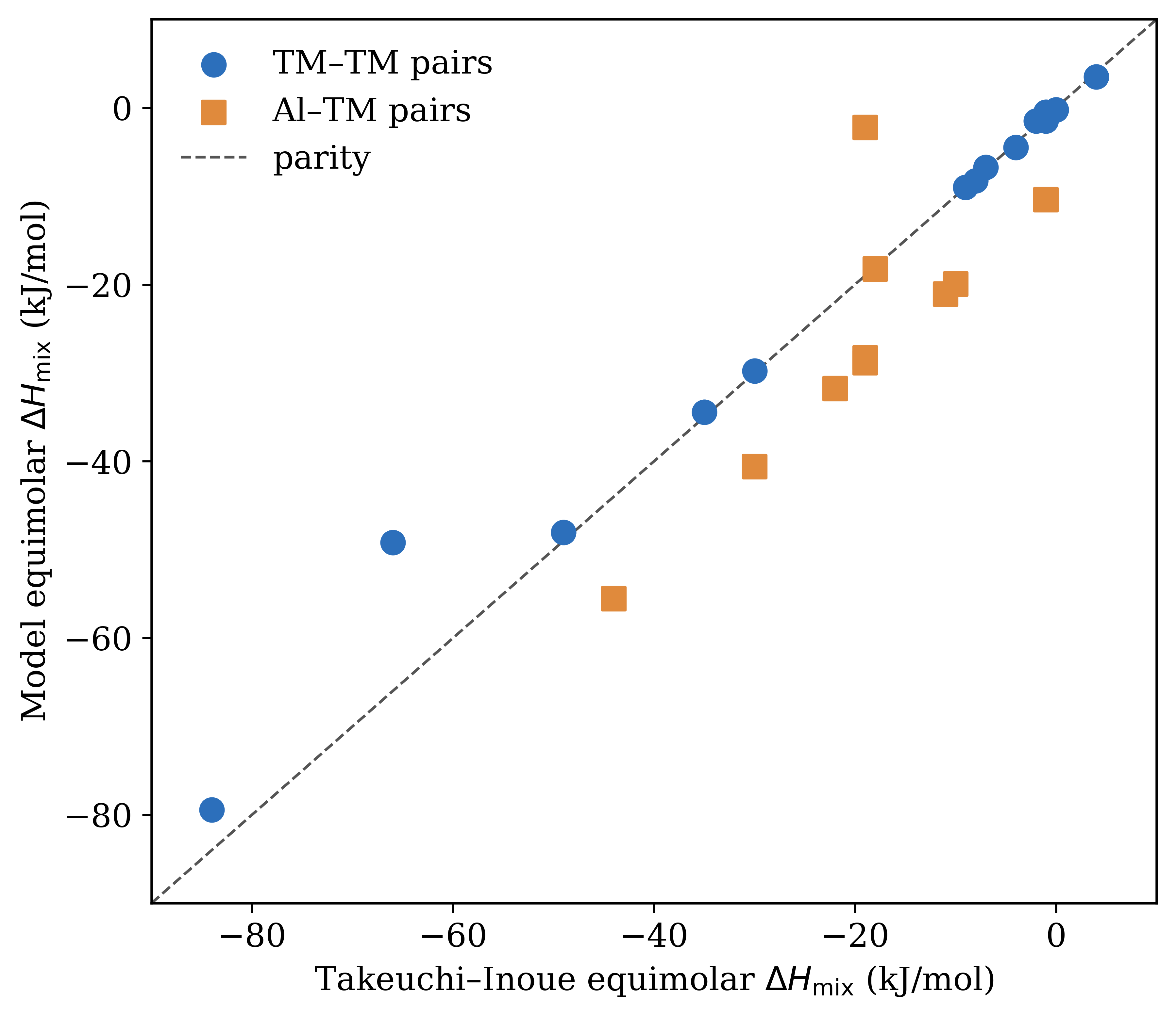}
		\caption{Surface-concentration-corrected macroscopic-atom-model equimolar binary enthalpies against the empirical Takeuchi--Inoue tabulation. Transition-metal pairs agree to within approximately ten percent; Al--TM pairs show the systematic Miedema overestimation expected for large electronegativity contrast. The dashed line is parity.}
		\label{fig:enthalpy}
	\end{figure}

	Before the structural and ordering terms are introduced, the chemical enthalpy that forms the common baseline of every candidate phase must itself be trustworthy, and Figure~\ref{fig:enthalpy} confirms that it is. The surface-concentration-corrected Miedema enthalpies of the TM binary pairs agree with the empirical Takeuchi--Inoue tabulation to within approximately ten percent, and the equimolar mixing enthalpy of the CoCrFeMnNi Cantor alloy is computed as $-4.18$~\si{kJ/mol} against the experimentally established value of $-4.3$~\si{kJ/mol}, an agreement within three percent. The Al--TM enthalpies are reproduced with the systematic overestimation characteristic of the Miedema model for pairs of large electronegativity contrast, a feature inherited from the underlying parametrization rather than introduced by the present treatment; because the structural and ordering terms that follow are independent of the absolute magnitude of the chemical enthalpy, the classifier is comparatively insensitive to this overestimation, and the agreement on the experimentally anchored Cantor composition demonstrates that the energetic backbone is sound.
	
	\subsection{The structural axis: electronic versus lattice-stability discrimination}
	
\begin{figure}[!ht]
		\centering
		\includegraphics[width=1\linewidth]{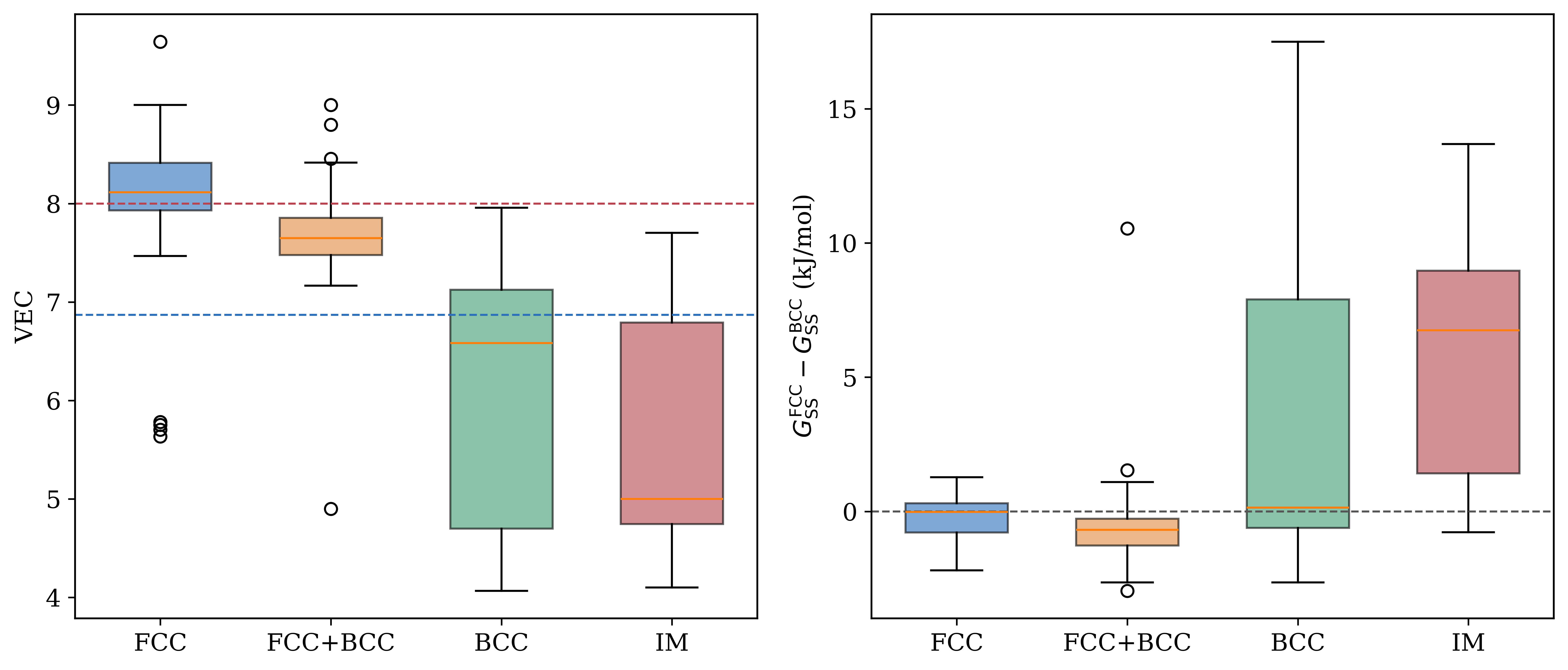}
		\caption{Discrimination of the experimental phase classes by (a) the valence-electron concentration, which orders the classes monotonically with the FCC/duplex and BCC boundaries shown as dashed lines, and (b) the free-energy structural gap from linearly mixed lattice stabilities, which fails to separate the duplex class from the single-phase FCC class. Boxes span the interquartile range; the central line is the median.}
		\label{fig:structural}
	\end{figure}

	A central design decision concerns which quantity should discriminate FCC from BCC stability, and Figure~\ref{fig:structural} resolves it empirically. When the experimentally characterized alloys are grouped by their reported phase, the valence-electron concentration orders the four classes monotonically, with the FCC alloys clustered at high VEC, the BCC and intermetallic alloys at low VEC, and the duplex alloys intermediate, the class means falling in the sequence $7.85$, $7.69$, $5.63$, and $5.57$ for FCC, duplex, BCC, and intermetallic respectively. The free-energy structural gap $G_{\mathrm{SS}}^{\mathrm{FCC}} - G_{\mathrm{SS}}^{\mathrm{BCC}}$ obtained from the linearly mixed lattice stabilities, by contrast, fails to order the classes: it does separate the clearly FCC alloys from the clearly BCC and intermetallic alloys, but it places the duplex class on the more strongly FCC-favoring side of the single-phase FCC class, an inversion that would cause a classifier relying on it to misassign the duplex alloys systematically. The origin of this failure is physical: the structural preference of a concentrated TM solution reflects the collective filling of the $d$-band, which the VEC captures directly but which a linear sum of pure-element lattice stabilities does not, because the pure-element penalties do not combine additively in the alloy. We therefore adopt the VEC as the structural discriminator and confine the free-energy treatment to the solid-solution-versus-intermetallic decision, a choice that is vindicated by the per-class performance reported below and that defines the scope of the contribution. The present method does not displace the electronic criterion for cubic structure selection but supplies the ordered-phase free energy that the scalar criteria lack.
	
	\subsection{The dominant-pair ordering mechanism}
	
\begin{figure}[!ht]
		\centering
		\includegraphics[width=1\linewidth]{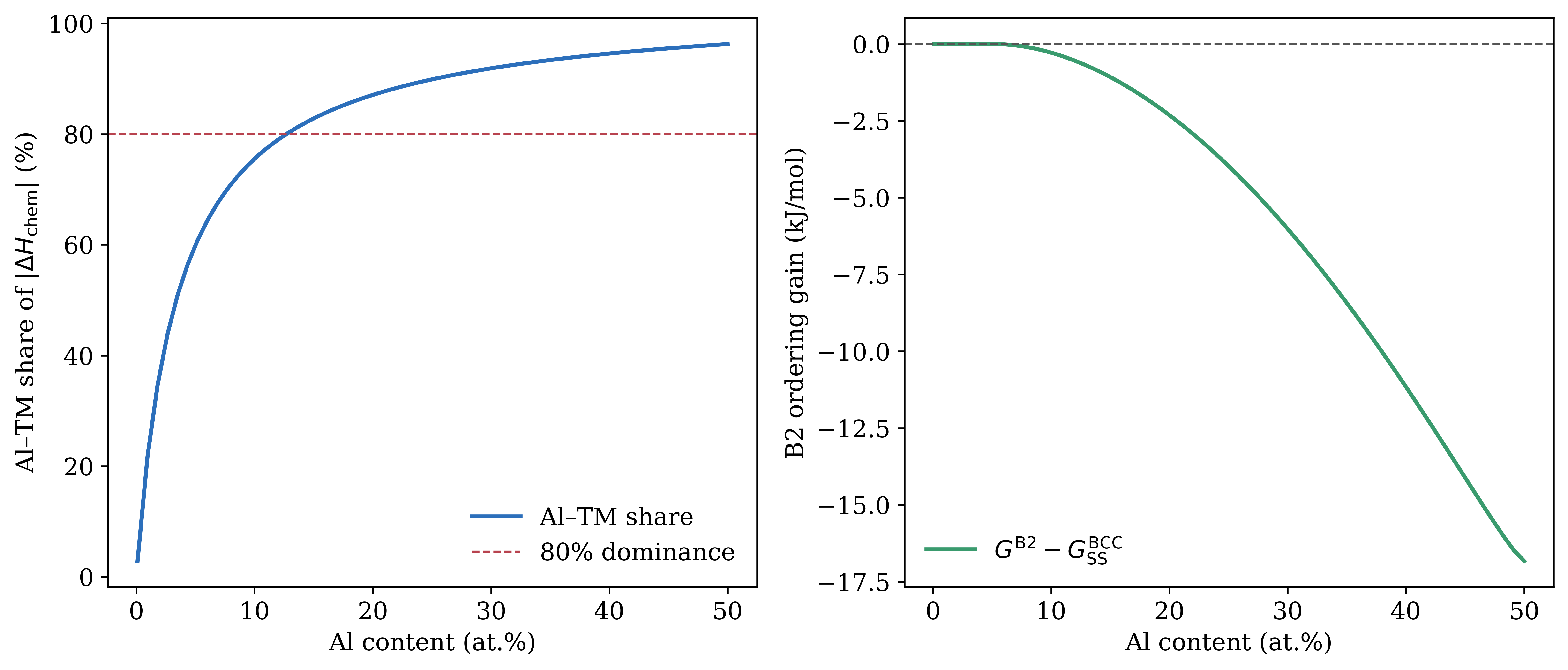}
		\caption{The dominant-pair mechanism for Al$_x$CoCrFeNi. (a) The Al--TM pairs contribute more than eighty percent of the chemical mixing enthalpy above approximately ten atomic percent Al, justifying the pseudo-binary reduction. (b) The resulting dominant-pair B2 ordering free energy falls increasingly below the disordered BCC solid solution as Al is added.}
		\label{fig:dominant}
	\end{figure}

	The ordered competitor that distinguishes the present model from the scalar criteria derives its tractability from the dominant-pair structure of the enthalpy, in which Figure~\ref{fig:dominant} documents for the ideal Al$_x$CoCrFeNi system. As Al is added, the fraction of the chemical mixing enthalpy contributed by the Al--TM pairs rises steeply, exceeding eighty percent above roughly ten atomic percent Al and approaching unity at higher contents, so that the enthalpy is governed by a single pair family rather than by the collective average of all pairwise interactions. This dominance is what justifies the reduction of the multicomponent B2-ordering problem to an effective pseudo-binary between Al and a composite TM species. The resulting ordering free energy of Equation~\eqref{eq:GB2} is shown in the second panel, where the B2 phase lies progressively below the disordered BCC solid solution as Al increases. The ordering gain grows in magnitude with composition, following the strengthening of the dominant interaction. The self-regulating critical temperature of Equation~\eqref{eq:Tc} ensures that this stabilization switches off smoothly where the dominant interaction is weak, so that the same construction that drives strong B2 ordering in the Al-rich alloys leaves the Al-free refractory alloys correctly described as disordered BCC solid solutions. The dominant-pair mechanism is thus not merely a computational convenience but the physical content of the model's distinctive predictive layer.
	
\begin{figure}[!ht]
		\centering
		\includegraphics[width=0.7\linewidth]{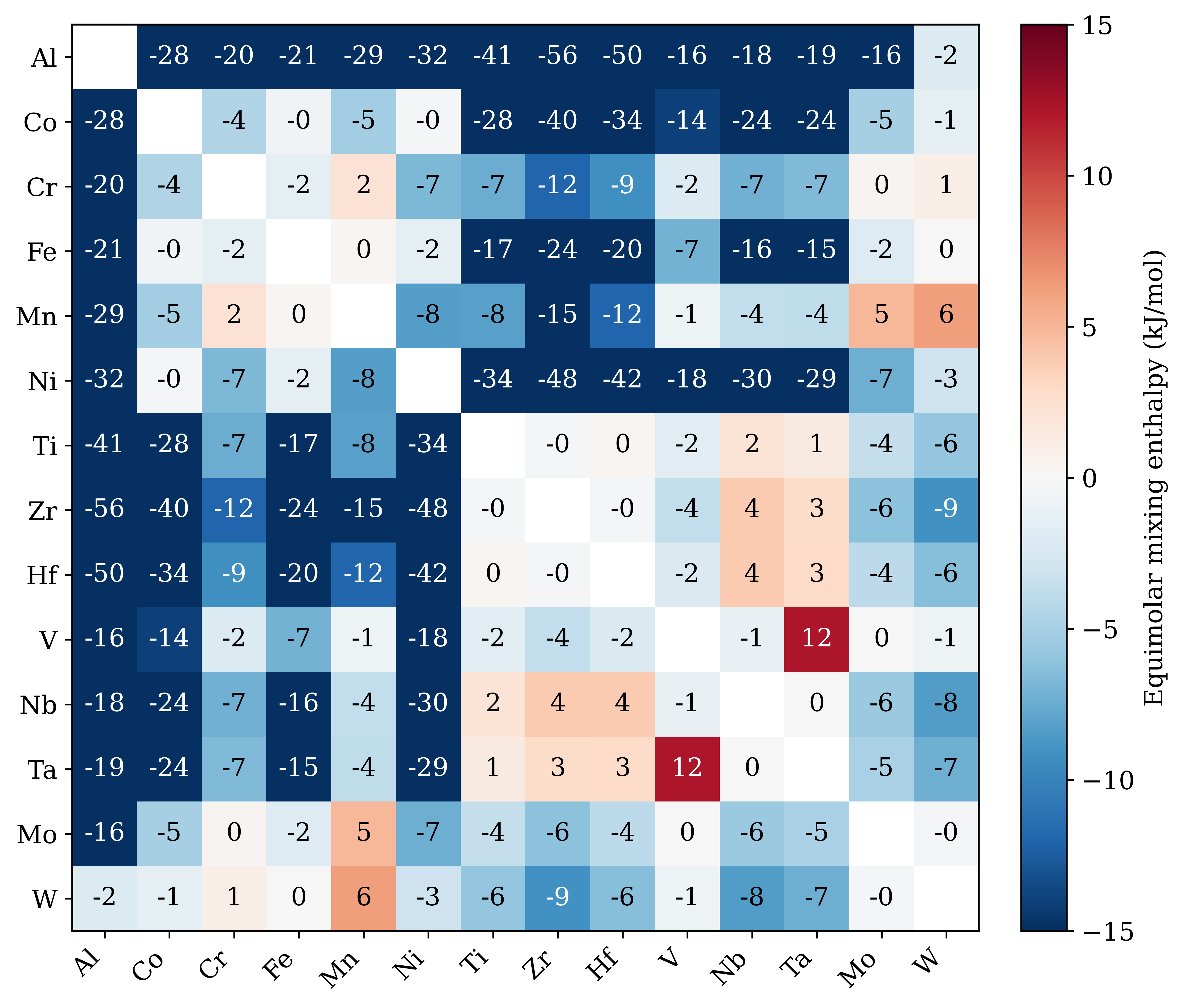}
		\caption{Equimolar pairwise mixing enthalpies for the fourteen principal elements. Al--TM pairs (top row and left column) are far stronger than the mutual TM interactions, concentrating the enthalpy into a small set of dominant pairs.}
		\label{fig:heatmap}
	\end{figure}

The dominance evident in Figure~\ref{fig:dominant} arises from the full interaction matrix shown in Figure~\ref{fig:heatmap}: the Al--TM entries are several times larger in magnitude than the mutual TM interactions, and the strongest single pairs such as Al with Ni, Co, and the early refractory metals stand out as the darkest cells. It is this concentration of the enthalpy into a few pairs that strengthens the pseudo-binary reduction of the ordering problem.
	
	\subsection{Validation of the dominant-pair ordering temperature}
	
\begin{figure}[!ht]
		\centering
		\includegraphics[width=1\linewidth]{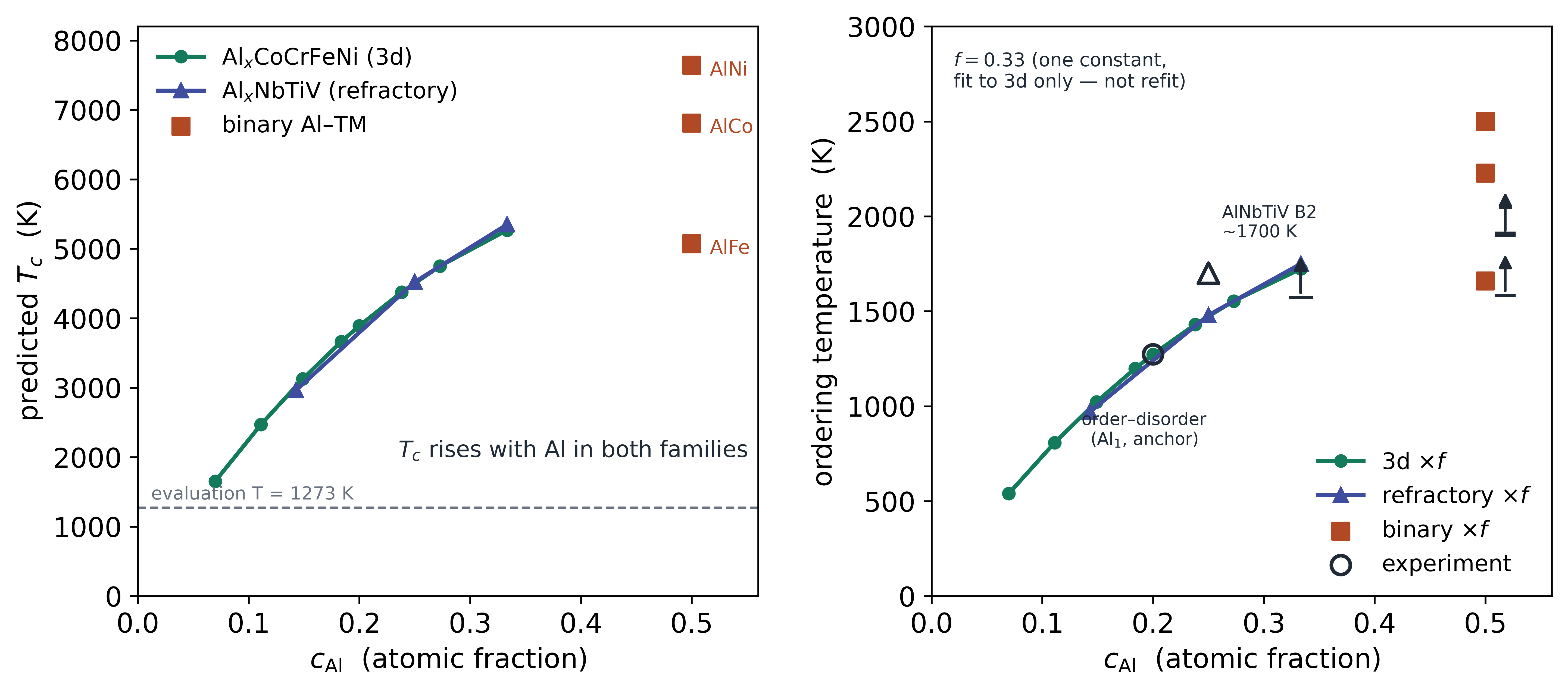}
		\caption{Validation of the dominant-pair ordering temperature of Equation~\eqref{eq:Tc}. (a) The predicted mean-field $T_c$ rises monotonically with Al content for both the $3d$ Al$_x$CoCrFeNi and the refractory Al$_x$NbTiV series, which collapse onto a common trend, and follows the rank order Al--Ni $>$ Al--Co $>$ Al--Fe for the binary aluminides; the dashed line marks the \SI{1273}{K} evaluation temperature. (b) A single composition-independent factor $f$, fitted only to the $3d$ order--disorder point and not refitted, brings the predicted ordering temperatures of both families into agreement with the experimental anchors (open symbols; upward arrows denote lower bounds for systems ordered to melting). The mean-field $T_c$ overestimates the absolute transition temperatures, but the trend, rank order, and cross-family transferability are parameter-free.}
		\label{fig:tcvalid}
	\end{figure}

	Equation~\eqref{eq:Tc} predicts a critical ordering temperature $T_c = -2\,\Omega_{\mathrm{RS}}\,x\,x_B/R$ set entirely by the strength of the dominant pair, and evaluating it directly from the model's surface-weighted interactions provides an independent, experimentally testable check on the ordering mechanism. Three parameter-free tests confirm that this temperature behaves physically across two chemically distinct families (Figure~\ref{fig:tcvalid}). First, in both the $3d$ Al$_x$CoCrFeNi and the refractory Al$_x$NbTiV series the predicted $T_c$ rises monotonically with Al content, and because the composition-weighted Al--TM interaction is comparable ($\approx -100$~kJ/mol) in both chemistries the two families collapse onto a common $T_c(c_{\mathrm{Al}})$ trend; this reproduces the experimentally established strengthening of B2 order with Al, for which AlNbTiV is a single-phase B2-ordered alloy~\cite{yurchenko2017alnbtivzr} with B2 ordering placed below $\approx\SI{1700}{K}$ by atomistic simulation~\cite{alnbtiv2021b2sim,woodgate2025jpm}. Second, for the binary aluminides, the predicted $T_c$ follows the order Al--Ni $>$ Al--Co $>$ Al--Fe, consistent with the retention of B2 order in NiAl and CoAl up to their melting points, whereas FeAl orders more weakly\cite{aluminide_phasediagrams}. Third, the construction self-regulates: Al-free refractory alloys lack a dominant B2-forming pair and are therefore correctly predicted to remain disordered; a Ti-bearing refractory alloy with a weak or repulsive interaction yields a $T_c$ far below the evaluation temperature; and only Al addition raises $T_c$ into the ordered regime, with the higher Al content required to order the refractory family (relative to the $3d$ family) arising directly from the weaker Al--refractory interactions. As expected for a mean-field treatment which neglects short-range order~\cite{woodgate2025jpm}, the absolute $T_c$ systematically overestimates the measured order--disorder temperatures; a single composition-independent scaling factor, fitted only to the $3d$ order--disorder point and not refitted, simultaneously places the refractory AlNbTiV ordering near its reported value and the binary aluminides above their melting points (Figure~\ref{fig:tcvalid}b). The mechanistic content, comprising the monotonic trend, the binary rank order, the cross-family transferability, and the self-regulation, requires no adjustment.
	
	\subsection{Mapping phase regions with the model}
	
\begin{figure}[!ht]
		\centering
		\includegraphics[width=0.7\linewidth]{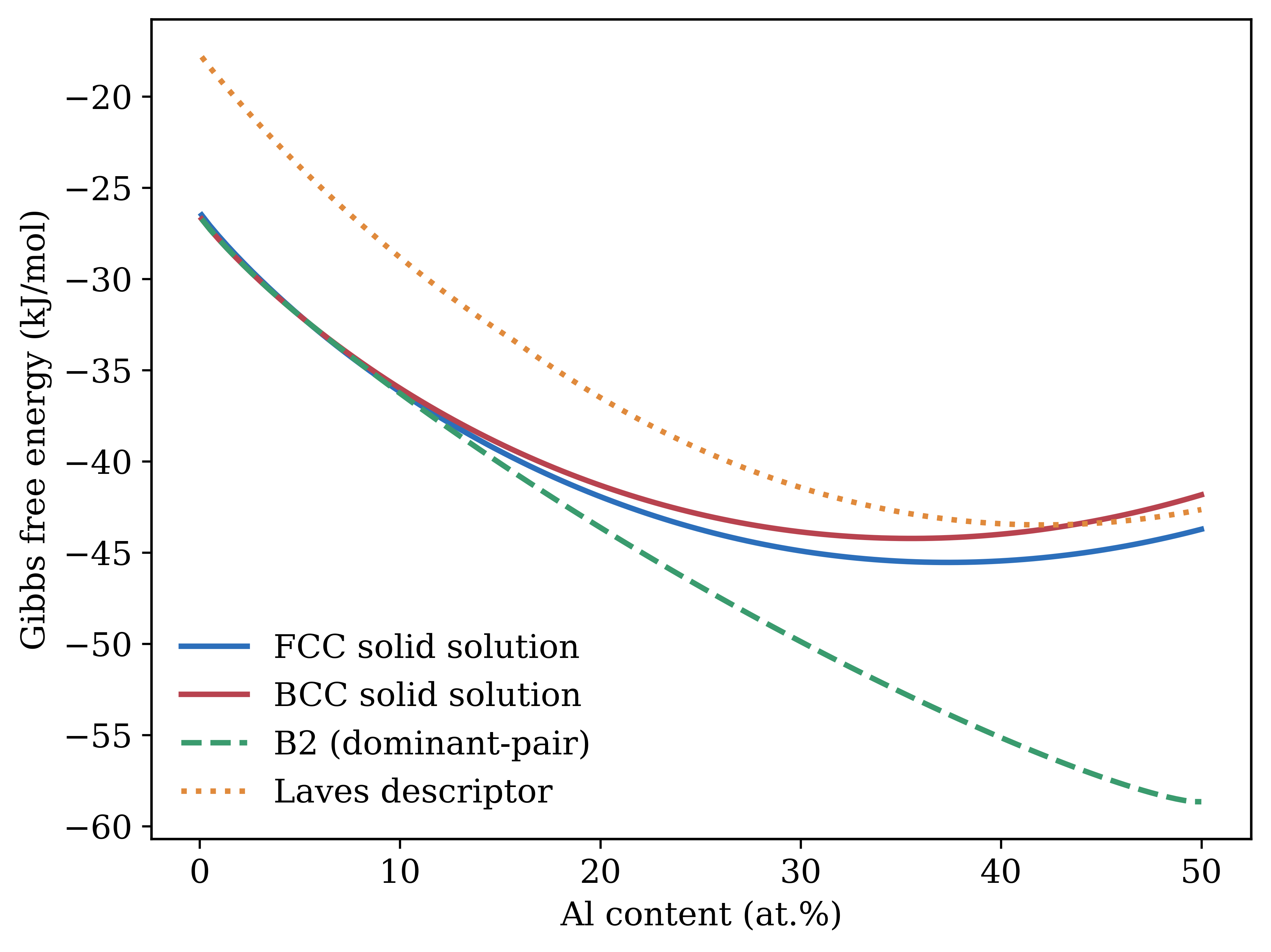}
		\caption{Free-energy competition along the Al$_x$CoCrFeNi trajectory at \SI{1273}{K}. The predicted phase sequence FCC $\rightarrow$ duplex $\rightarrow$ BCC/B2 arises from the crossing of the candidate free-energy curves, the dominant-pair B2 phase descending below the solid solutions as Al increases.}
		\label{fig:crossover}
	\end{figure}

\begin{figure}[!ht]
		\centering
		\includegraphics[width=1\linewidth]{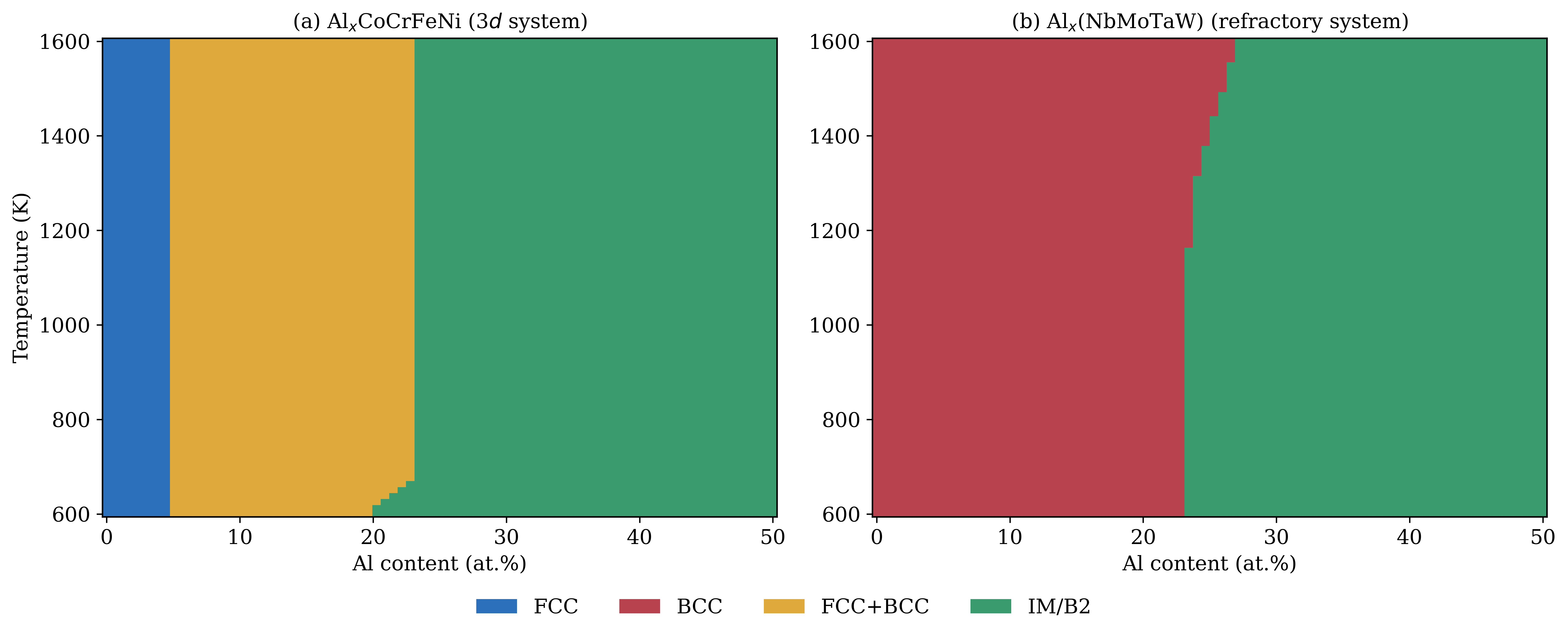}
		\caption{Predicted temperature--composition phase maps for (a) the $3d$ system Al$_x$CoCrFeNi and (b) the refractory system Al$_x$(NbMoTaW). In the $3d$ system a single-phase FCC field at low Al gives way to a duplex band and then to ordered B2/intermetallic; in the refractory system a single-phase BCC field persists to higher Al before ordering takes over. In both, the ordered-field boundary advances to lower Al as temperature falls, through the entropic and ordering terms of the free energy.}
		\label{fig:Tx}
	\end{figure}

\begin{figure}[!ht]
		\centering
		\includegraphics[width=0.7\linewidth]{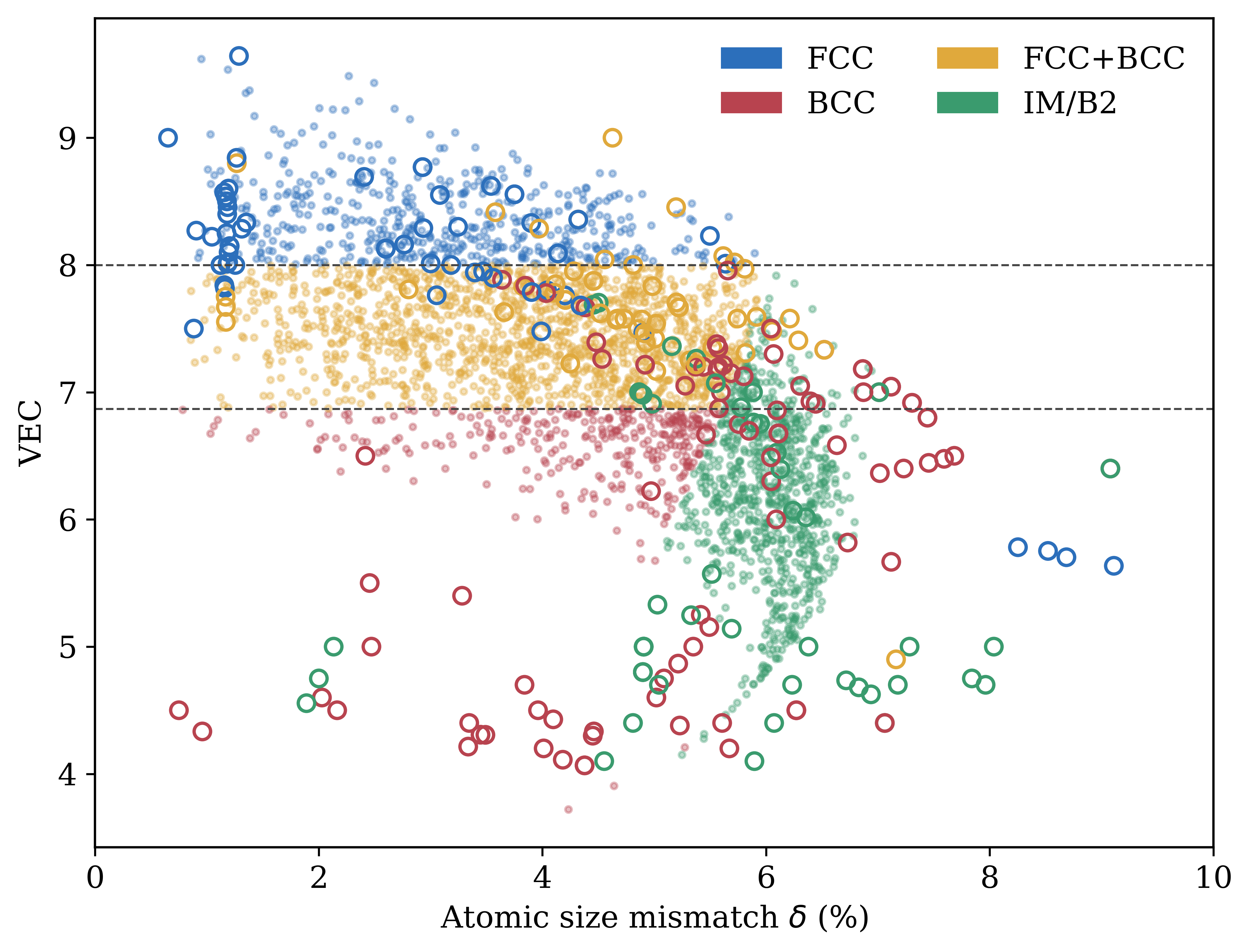}
		\caption{Model-predicted phase map of the $3d$+Al composition space. Filled points are several thousand sampled compositions colored by predicted phase; open rings are the experimentally characterized alloys at their reported labels. The predicted phases occupy coherent regions, and the experimental alloys fall predominantly within them.}
		\label{fig:phasemap}
	\end{figure}

	Because the classifier returns a phase for any composition and temperature, it can be used not only to label individual alloys but to map continuous regions of the design space, a capability the scalar criteria share only in projection and which Figures~\ref{fig:crossover}--\ref{fig:phasemap} exploit. Figure~\ref{fig:crossover} traces the four candidate free energies along the Al$_x$CoCrFeNi trajectory at \SI{1273}{K}. The FCC solid solution is lowest at low Al, the B2 ordered phase descends below both solid solutions as Al increases, and the predicted sequence FCC $\rightarrow$ duplex $\rightarrow$ BCC/B2 emerges directly from the crossing of these curves rather than from any imposed threshold. Resolving this competition over temperature as well as composition yields the predicted phase maps of Figure~\ref{fig:Tx}, which are, to our knowledge, a form of output the parametric criteria cannot produce. 
	
	For the $3d$ system Al$_x$CoCrFeNi (panel a), a single-phase FCC field at low Al gives way to a broad duplex band and then to a B2 and intermetallic field at high Al; for the refractory system Al$_x$(NbMoTaW) (panel b), a single-phase BCC field instead persists to higher Al before ordering takes over. The contrast between the two panels reflects the opposite VEC regimes of the $3d$ and refractory families. In both, the boundary of the ordered field tilts toward lower Al as temperature falls, reproducing the experimentally familiar tendency of as-cast single-phase alloys to develop ordered and secondary phases on cooling or annealing. This temperature dependence enters through the entropic terms of Equation~\eqref{eq:GSS} and, in the ordered phase, through the self-regulating critical temperature of Equation~\eqref{eq:Tc}, so that the maps are a genuine prediction of the model's free-energy balance rather than a fitted boundary; the same machinery is what allows each alloy in the benchmark to be evaluated at its own recorded characterization temperature. 
	
	Finally, Figure~\ref{fig:phasemap} maps the model's predictions across the full $3d$+Al composition space, sampling several thousand compositions and projecting them onto the VEC and size-mismatch plane. The predicted phases occupy coherent, contiguous regions, with FCC at high VEC and low size mismatch, BCC and intermetallic phases at low VEC, and duplex regions between them. The experimentally characterized alloys, overlaid as rings, fall predominantly within the corresponding predicted regions. The agreement is a visual counterpart to the quantitative benchmark that follows. Because these free energies are built from elemental inputs tabulated for essentially every element, the construction is not specific to high-entropy compositions. The same minimum-free-energy comparison can be carried out for any multicomponent alloy at any temperature. The HEAs mapped here are therefore a demonstration of the method rather than a limit on it, and conventional or dilute multicomponent systems can be screened with exactly the same machinery.

	\subsection{Four-class benchmark against the parametric criteria}
	
\begin{figure}[!ht]
		\centering
		\includegraphics[width=0.8\linewidth]{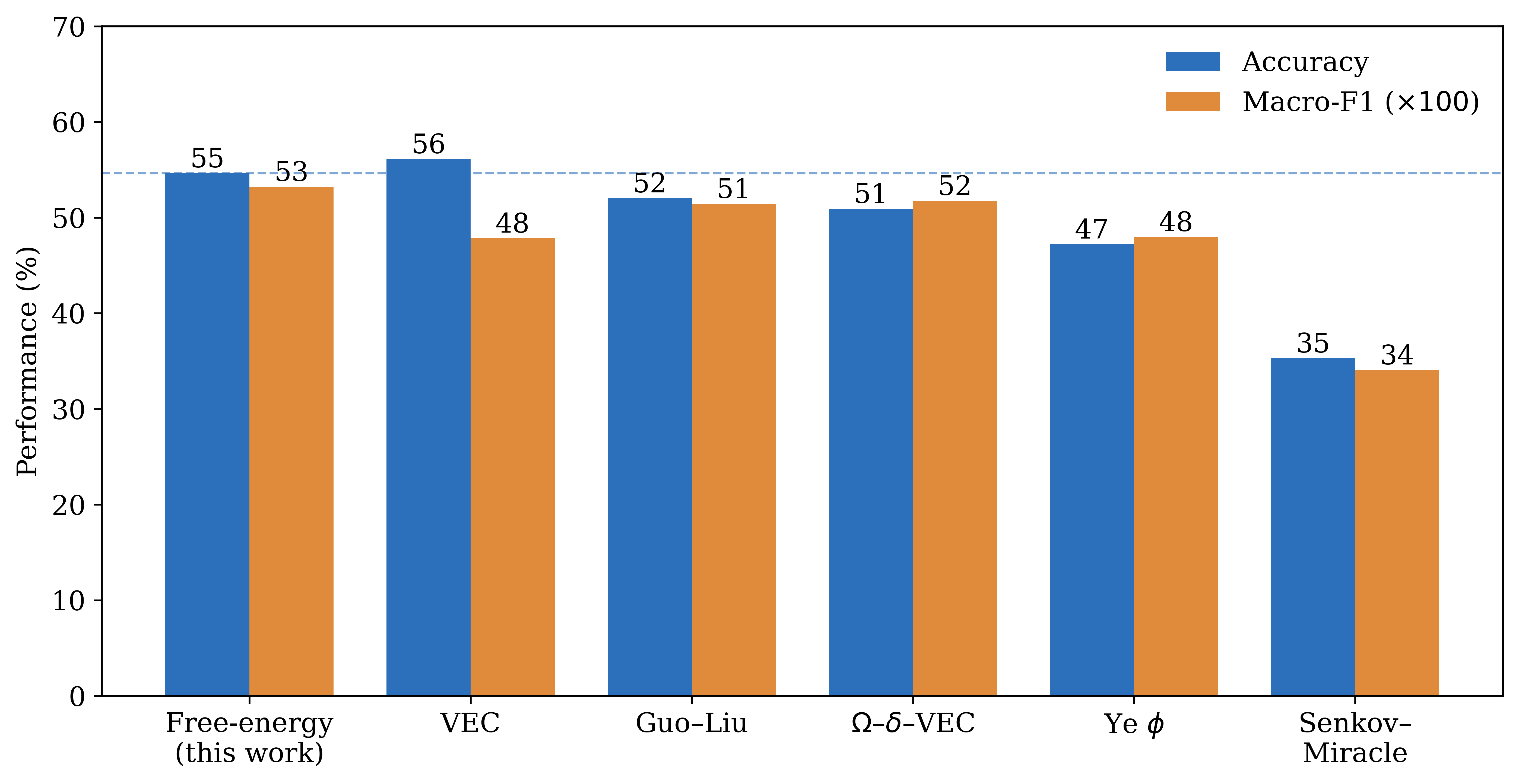}
		\caption{Four-class phase-prediction performance of the free-energy classifier and the five parametric criteria on $269$ alloys under five-fold cross-validation, by accuracy and macro-F1. The free-energy classifier leads on the class-balanced macro-F1; the valence-electron criterion attains marginally higher raw accuracy but falls well short on macro-F1 because it has no mechanism for the intermetallic class. The dashed line marks the free-energy classifier's macro-F1.}
		\label{fig:comparison}
	\end{figure}

	The principal comparison is presented in Figure~\ref{fig:comparison} and Table~\ref{tab:fourclass}, which report the cross-validated four-class performance of the free-energy classifier and the five parametric criteria on the $269$ in-scope alloys. The free-energy classifier attains the highest macro-F1 of any method, $0.532$, at an accuracy of $54.6\%$. The interpretation of these figures requires care, because the VEC criterion attains a slightly higher raw accuracy of $56.1\%$ while falling well short on the macro-F1 at $0.478$. This pattern is diagnostic rather than incidental because the VEC criterion achieves its accuracy by performing well on the three solid-solution classes, which constitute the majority of the dataset. At the same time it possess no mechanism whatever for predicting the intermetallic class, on which it therefore fails entirely. Its raw accuracy is thus purchased by neglecting the hardest class, precisely the behavior that the macro-F1, which weights all four classes equally, is designed to penalize; for an imbalanced four-class proble, the macro-F1 is the faithful measure of predictive quality, and on it the free-energy classifier is the strongest of the six. The free-energy classifier's advantage on the macro-F1 reflects its balanced performance across all four phases, including the intermetallic class that the electronic criterion cannot address. Against the enthalpy-based criteria the advantage is unambiguous on both metrics: the Guo--Liu, $\Omega$--$\delta$--VEC, and Ye $\phi$ criteria cluster between forty-seven and fifty-two percent accuracy with macro-F1 scores between $0.48$ and $0.52$, and the Senkov--Miracle $k_1^{\mathrm{cr}}$ criterion, evaluated with its exact published formula at the characterization temperatures, trails the field at $35.3\%$. That the $\phi$ and $k_1^{\mathrm{cr}}$ baselines are implemented faithfully where the former verified against its reported reference values ensures that the comparison reflects the genuine performance of these criteria rather than any weakening in their reproduction.
	\begin{table}[!ht]
		\centering
		\caption{Four-class phase prediction (FCC/BCC/FCC$+$BCC/intermetallic) across $269$ alloys, evaluated by five-fold cross-validation. The free-energy classifier attains the highest macro-F1; the VEC criterion attains marginally higher raw accuracy but the lowest intermetallic recall.}
		\label{tab:fourclass}
		\begin{tabular}{lcc}
			\toprule
			Method & Accuracy & Macro-F1 \\
			\midrule
			Free-energy classifier (this work) & 54.6\% & 0.532 \\
			Valence-electron concentration & 56.1\% & 0.478 \\
			Guo--Liu $\Delta H$--$\delta$ & 52.0\% & 0.514 \\
			Yang--Zhang $\Omega$--$\delta$--VEC & 50.9\% & 0.518 \\
			Ye $\phi$ & 47.2\% & 0.480 \\
			Senkov--Miracle $k_1^{\mathrm{cr}}$ & 35.3\% & 0.341 \\
			\bottomrule
		\end{tabular}
	\end{table}
	
\begin{figure}[!ht]
		\centering
		\includegraphics[width=1\linewidth]{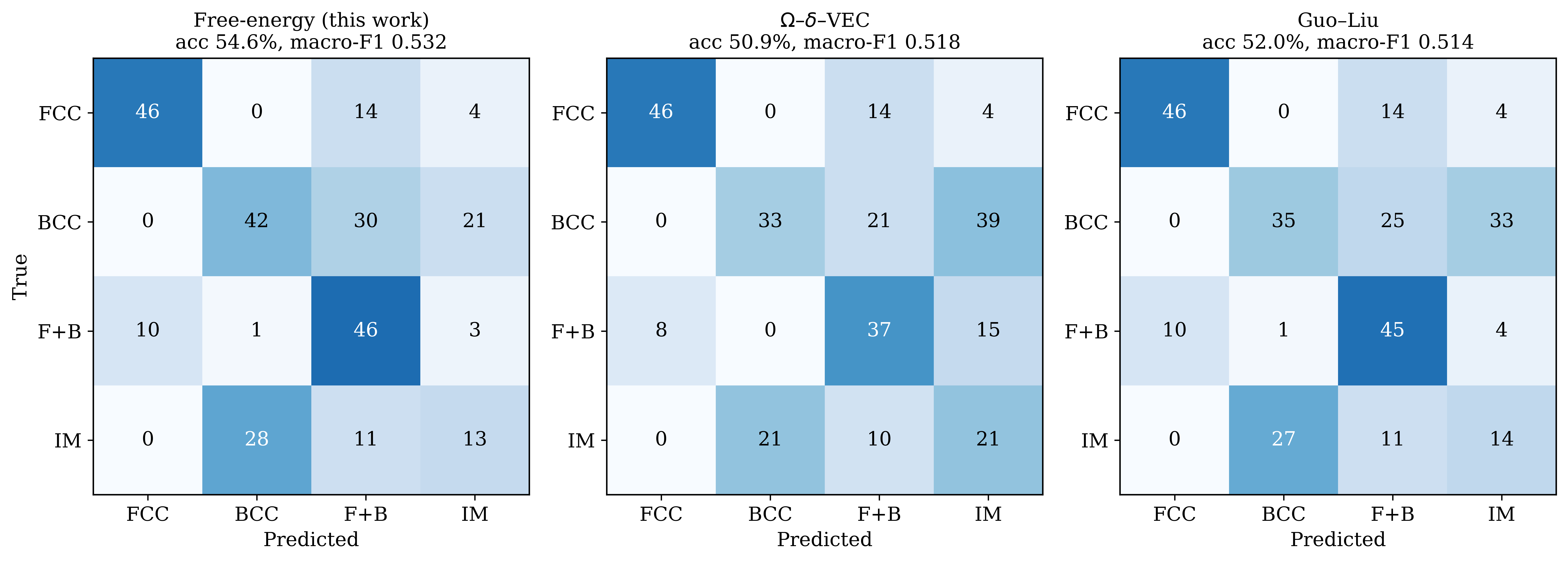}
		\caption{Four-class confusion matrices for the free-energy classifier and two representative parametric criteria. Rows are the experimentally reported phases and columns the predictions; cell entries are alloy counts and the colour encodes the row-normalized rate. The parametric criteria over-assign the intermetallic class by rejecting high-size-mismatch refractory alloys, whereas the free-energy classifier recovers many genuine BCC and duplex alloys.}
		\label{fig:confusion}
	\end{figure}

	The confusion structure underlying these aggregate figures, shown in Figure~\ref{fig:confusion}, locates the origin of the free-energy classifier's advantage. The $\Omega$--$\delta$--VEC criterion misclassifies a large number of genuine BCC and duplex alloys as intermetallic, because its requirement that the atomic-size mismatch fall below $6.6\%$ indiscriminately rejects the high-size-mismatch refractory alloys, many of which are in fact stable single-phase or duplex solid solutions whose large size mismatch is compensated by their high melting temperatures. The free-energy classifier, by computing an explicit ordering free energy through the dominant-pair B2 model and a size-aware Laves descriptor, distinguishes the alloys in which ordering is genuinely strong from those in which a large size mismatch coexists with a stable disordered solution, recovering many of the BCC and duplex alloys that the size-mismatch cutoff discards. The improvement is therefore concentrated precisely on the solid-solution-versus-intermetallic decision, which is the axis that the present model adds to the electronic structural discrimination and which the scalar criteria handle only through blunt thresholds.
	
\begin{figure}[!ht]
		\centering
		\includegraphics[width=1\linewidth]{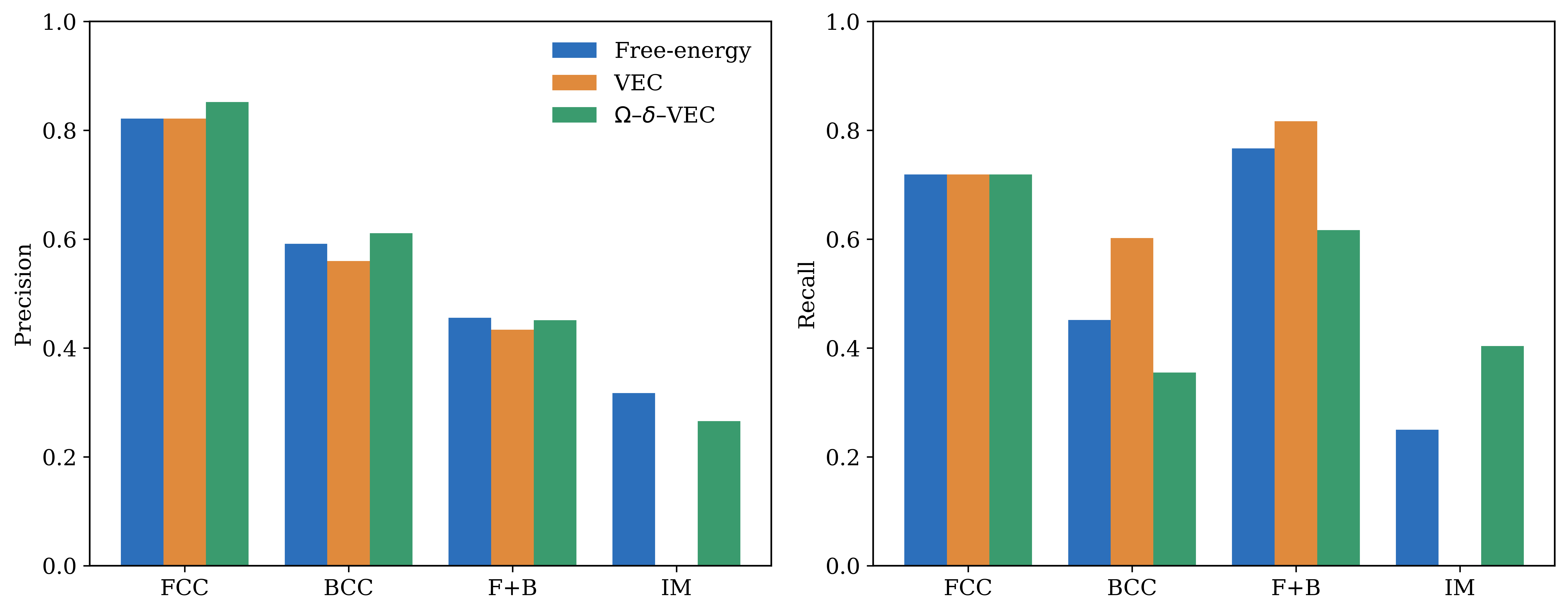}
		\caption{Per-class (a) precision and (b) recall for the free-energy classifier and two representative parametric criteria. The free-energy classifier leads on precision for the duplex and intermetallic classes, and is alone in recording non-zero recall on the intermetallic class, where the valence-electron criterion fails entirely. No single method dominates every class; the free-energy classifier's advantage is its balance across all four, which is what the class-balanced macro-F1 rewards.}
		\label{fig:perclass}
	\end{figure}

	The class-resolved view of Figure~\ref{fig:perclass} shows that the macro-F1 advantage arises from balance rather than from dominance of any single class where the free-energy classifier leads on duplex and intermetallic precision and is alone in recording non-zero intermetallic recall, while no method is best on every class.
	
	Resolved by class, the free-energy classifier's held-out performance makes the same point quantitatively. It attains the highest precision on the duplex class, at $0.46$, and the highest precision on the intermetallic class, at $0.32$, the latter against zero for the valence-electron criterion, which assigns no alloy to the intermetallic class at all. Its FCC performance matches the parametric criteria (F1 $0.77$), reflecting the clean separation of high-valence-electron compositions, and its duplex recall, at $0.77$, trails only the valence-electron band. No single method dominates every class: the $\Omega$--$\delta$--VEC criterion achieves higher intermetallic recall and the valence-electron band higher duplex recall, but each does so at the cost of a class on which it collapses---the valence-electron criterion to zero intermetallic recall, the $\Omega$--$\delta$--VEC criterion to the lowest BCC recall of the three. The free-energy classifier alone avoids any such collapse, scoring non-trivially on all four classes, and it is this balance rather than the dominance of any single class that yields the highest macro-F1 and that the metric is designed to reward. The intermetallic class remains the weakest for every method, a signature of the resolution ceiling examined next, since a substantial part of that class is, as we show, not recoverable by any equilibrium primary-phase model.
	
	\subsection{A resolution ceiling intrinsic to four-class prediction}
	
\begin{figure}[!ht]
		\centering
		\includegraphics[width=0.7\linewidth]{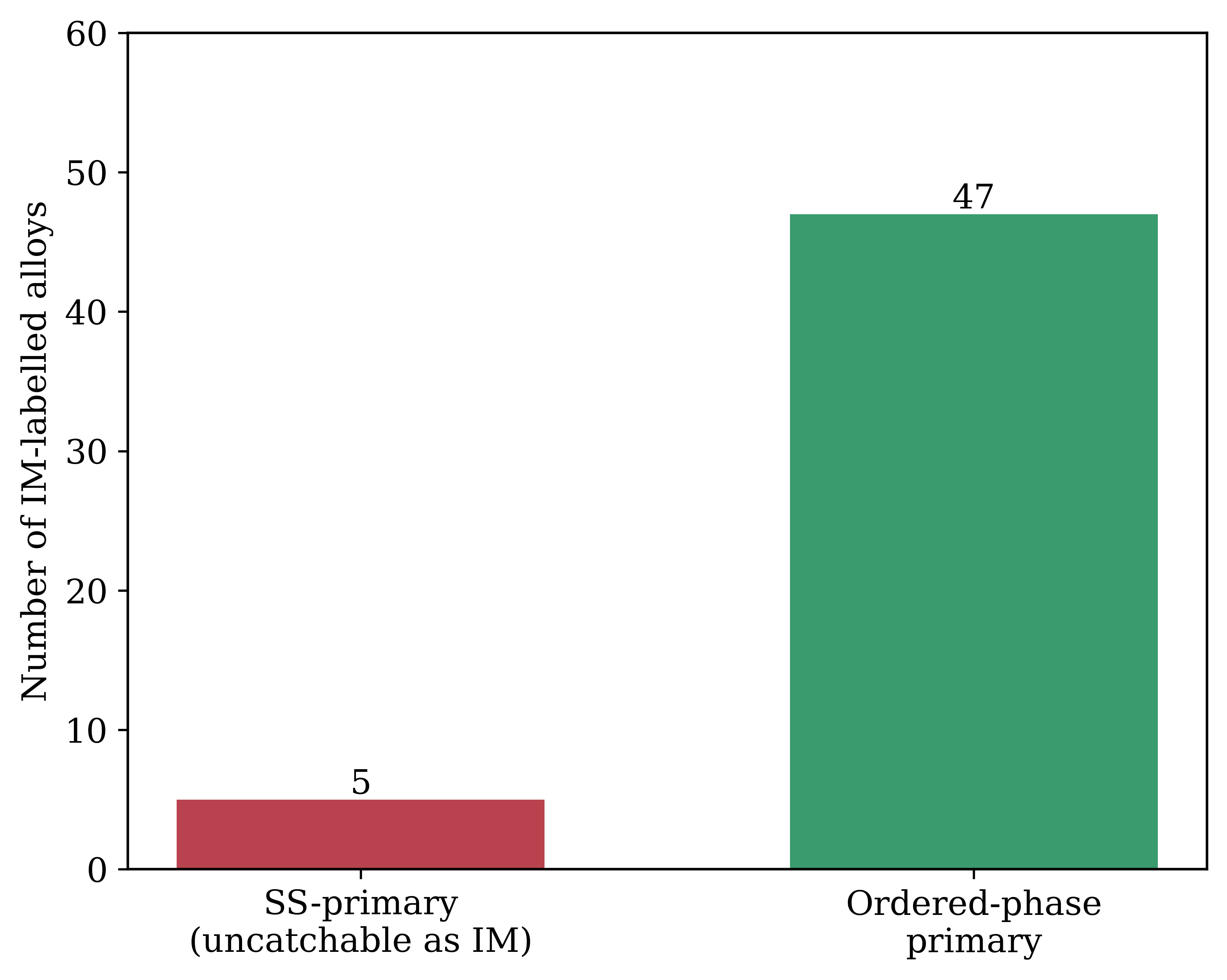}
		\caption{The phase-label resolution ceiling. A subset of the alloys labeled intermetallic in the literature possess a solid solution as their thermodynamically dominant phase, the intermetallic label denoting only a minor secondary constituent; no equilibrium primary-phase model can assign these the intermetallic label, which sets a ceiling on four-class accuracy that limits all methods. The exact count is parametrization-dependent while the ceiling's existence is robust.}
		\label{fig:ceiling}
	\end{figure}

	The four-class accuracies of all six methods plateau between thirty-five and fifty-six percent, and Figure~\ref{fig:ceiling} demonstrates that this plateau is not a deficiency of any particular model but a ceiling imposed by the structure of the prediction task itself. Of the fifty-two alloys labeled intermetallic or multiphase in the dataset, five have a solid solution as their thermodynamically lowest single phase. These are microstructures in which a solid-solution matrix coexists with a minor secondary intermetallic, so that the alloy's primary or matrix phase is genuinely a solid solution and the intermetallic label reports only the existence of a secondary constituent rather than the identity of the dominant phase. No equilibrium model that determines the most stable single phase, including the present classifier and the parametric criteria, can correctly assign the intermetallic label to such an alloy because the most stable phase is, by construction, the solid-solution matrix. We note that the precise count is mildly sensitive to the lattice-stability parametrization where a further group of low-Al, $3d$-rich alloys sits within approximately $1$~kJ/mol of the solid-solution--to--B2 crossover, so that small changes in the structural energies move them between a solid-solution-primary and an ordered-primary assignment; their near-degeneracy is itself consistent with their being reported experimentally as multiphase. The existence of the ceiling is therefore robust even though its exact magnitude is parametrization-dependent. The literature category of intermetallic or multiphase conflates two physically distinct situations: an intermetallic as the primary phase and a trace intermetallic coexisting with a solid-solution matrix. This conflation places a ceiling on the achievable accuracy of four-class prediction that limits every method evaluated here. Recognizing the ceiling reframes the interpretation of the four-class figures: a portion of the residual error reflects not inadequacy of the thermodynamic models but an ill-posedness in the prediction target as recorded in the experimental literature, and the appropriate response is to evaluate the models on a target that is well-defined.
	
	\subsection{The well-posed three-class problem}
	
\begin{figure}[!ht]
		\centering
		\includegraphics[width=1\linewidth]{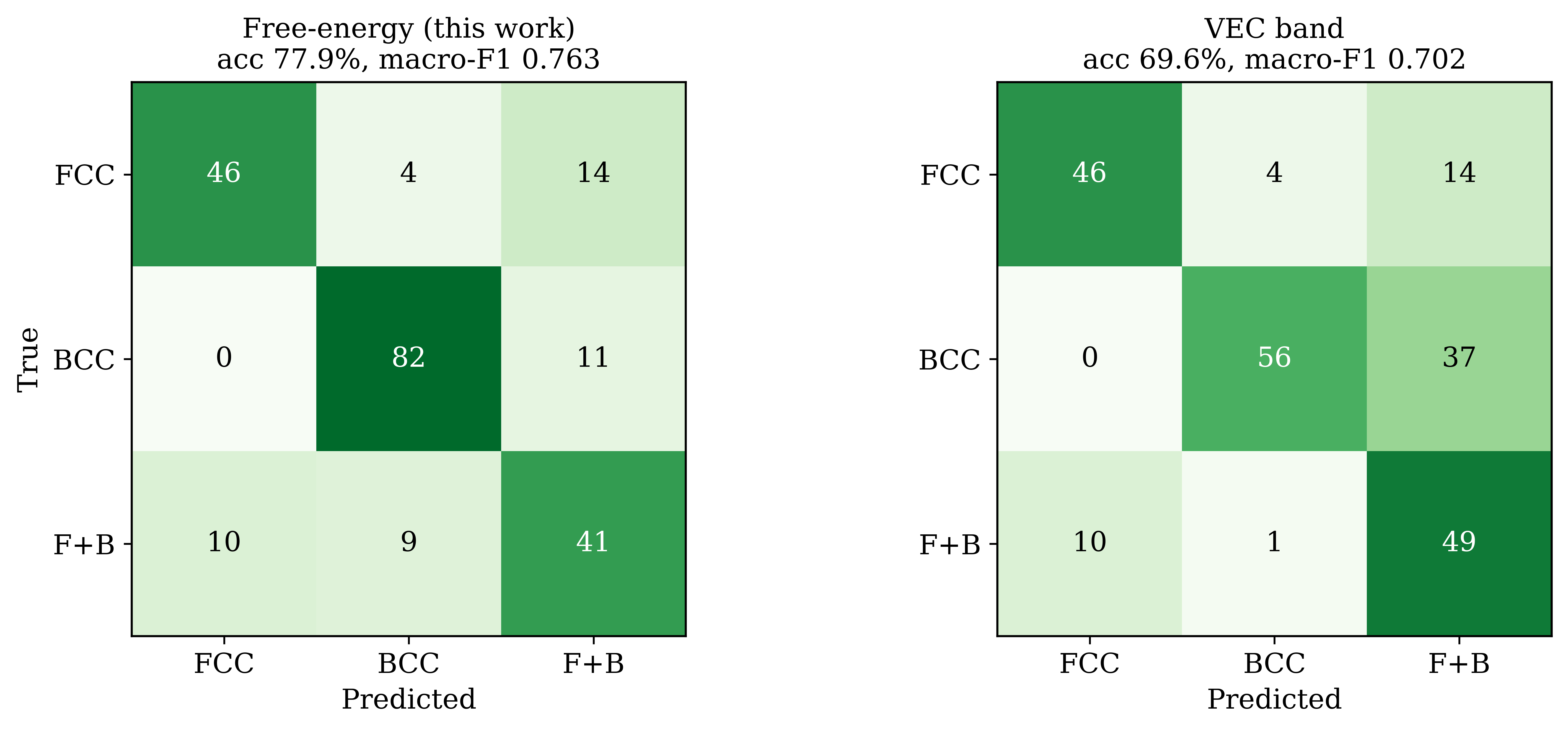}
		\caption{Three-class phase prediction (FCC/BCC/FCC$+$BCC) on the $217$ directly-labeled alloys, by five-fold cross-validation. The free-energy classifier reaches $77.9\%$ accuracy against $69.6\%$ for the valence-electron band, the advantage concentrated in the duplex class. Intermetallic-labeled alloys are excluded rather than relabeled, so the comparison is non-circular.}
		\label{fig:threeclass}
	\end{figure}

	Removing the ill-defined intermetallic class and evaluating the models on the $217$ alloys carrying a directly reported FCC, BCC, or duplex label yields the well-posed three-class problem of Figure~\ref{fig:threeclass}, on which the free-energy classifier reaches $77.9\%$ accuracy and a macro-F1 of $0.763$, against $69.6\%$ and $0.702$ for the valence-electron criterion. The intermetallic-labeled alloys are dropped from this evaluation rather than reassigned by the model's own energy, so that the three-class comparison is free of any circularity that would arise from relabeling alloys using the same free energy against which the model is then scored. Two features of this result are significant. First, the margin over the electronic criterion widens from the fraction of a percentage point seen on raw four-class accuracy to roughly eight percentage points, demonstrating that on the task it can actually decide, the free-energy treatment is substantially more powerful than the electronic threshold alone. Second, the gain is concentrated in the duplex FCC$+$BCC class, the class on which the structural discrimination is most demanding, indicating that the explicit free-energy comparison resolves the FCC-to-BCC crossover and its intervening two-phase window more faithfully than a single electronic boundary. The three-class result is the appropriate headline measure of the model's predictive value, because it reports performance on the question that equilibrium thermodynamics is well-posed to answer, and on that question the advantage over the established criteria is decisive.
	
	\subsection{Scope and limitations}
	
	Three limitations bound the present results and delineate the scope of the contribution. The first concerns the structural axis because the linearly mixed lattice stabilities do not reproduce the experimental FCC/BCC/duplex ordering as faithfully as the electronic VEC. The model adopts the latter for cubic structure discrimination, and the framework is consequently parameter-light rather than strictly parameter-free. Its novelty lies in the explicit free-energy treatment of the solid-solution-versus-intermetallic decision through the dominant-pair B2 model, not in displacing the electronic criterion for FCC-versus-BCC selection, and the contribution should be understood in those terms. The second limitation concerns the Laves competitor, which is represented by an approximate size-driven descriptor rather than a rigorous two-sublattice compound-energy model, because the latter would require intermetallic end-member energies from density-functional or CALPHAD sources that the present work deliberately excludes in order to keep the model self-contained; intermetallic recall accordingly remains the weakest element of the four-class performance, and a first-principles Laves free energy is the most promising route to improving it. The third concerns the lattice stabilities. These are taken from the SGTE unary database at a single representative temperature rather than as the full temperature-dependent functions evaluated at each alloy's individual characterization temperature. Because the classifier was found to be insensitive to temperature within the annealing regime, this is expected to be a minor effect, but evaluating the temperature-dependent functions per alloy is a natural and available refinement. A manual inspection of the misclassified alloys was carried out to determine whether any reflected errors in the curated labels rather than genuine limitations of the model; no mislabeled entries were identified, the misclassifications being dominated instead by alloys lying near the FCC/BCC boundary and by the solid-solution-matrix plus minor-intermetallic alloys responsible for the resolution ceiling, so no alloys were removed on the basis of being misclassified. None of these limitations directly or indirectly alters the central conclusions, namely that a parameter-light free-energy model out-predicts the standard parametric criteria, that four-class prediction is bounded by a quantified resolution ceiling that limits all methods, and that on the well-posed three-class problem the model's advantage is decisive.
	
	\section{Conclusions}
	
	We have shown that the ordered-phase free energy missing from every scalar phase-selection criterion can be reconstructed from tabulated elemental data alone, without density-functional or CALPHAD input, through a dominant-pair mechanism in which a single Al-TM interaction family supplies most of the mixing enthalpy in Al-bearing alloys. This collapses the intractable multicomponent B2-ordering problem onto an effective pseudo-binary whose Bragg-Williams free energy is computed exactly, providing the explicit ordered competitor structurally absent from the parametric criteria and self-regulating so that Al-free refractory alloys remain correctly disordered. The central novelty is that the framework assigns an explicit free energy to every candidate state rather than compressing the decision into one fitted scalar. Because it ranks a free energy for each competitor, the classifier predicts the lowest-energy phase at any composition and temperature, mapping continuous phase regions across the whole design space, an output the parametric criteria cannot produce. The model wins the class-balanced macro-F1 of six methods and reaches 77.9\% on the well-posed three-class task, eight points above VEC, while four-class accuracy is bounded by an ill-posedness in the experimental labels that limits every method. Drawing only on tabulated elemental inputs, the framework extends unchanged to any multicomponent alloy and temperature.

	\section*{Acknowledgments}
	This research was supported by the NSERC Alliance International Catalyst (ALLRP 592696-24), Canada, and the use of a high-performance computing system at the University of Manitoba and the Research Alliances of Canada. 
	
	\section*{CRediT authorship contribution statement}
	\textbf{Dennis Boakye}: Writing – review and editing, Writing – original draft, Visualization, Validation, Methodology, Investigation, Formal analysis, Data curation. \textbf{Chuang Deng}: Writing – review and editing, supervision, software, resources, project administration, investigation, Fund Acquisition, conceptualization.
	
	\section*{Declarations}
	The authors declare that they have no known competing financial interests or personal relationships that could have influenced the work reported in this paper.
	
	\section*{Data availability}
	The curated phase dataset, the Miedema interaction database, and the classifier source code are available from the authors on reasonable request.
	
	\section*{Supplementary information}
	The supplementary information referenced in the main text is attached as the Supplemental material.
	
	\scriptsize
	\bibliographystyle{elsarticle-num} 
	\biboptions{sort&compress}
	\bibliography{references}
	
\end{document}